\documentclass[traditabstract]{aa}
\bibpunct{(}{)}{;}{a}{}{,} 
\usepackage{graphicx}
\usepackage[varg]{txfonts}
\usepackage{color}
\begin{document}

\title{FeI and NiI in cometary atmospheres}
\subtitle{Connections between the NiI/FeI abundance ratio and chemical characteristics of Jupiter-family and Oort-cloud comets\thanks{Based on observations made with the ESO Very Large Telescope at the Paranal Observatory under program ID~102.C-0438.}}
\author{D. Hutsem\'ekers\inst{1},
        J. Manfroid\inst{1},
        E. Jehin\inst{1},
        C. Opitom\inst{2},
        Y. Moulane\inst{1,3}
        }
\institute{
    Institut d'Astrophysique et de G\'eophysique, Universit\'e de Li\`ege, All\'ee du 6 Ao\^ut 19c, 4000 Li\`ege, Belgium
    \and
    Institute for Astronomy, University of Edinburgh, Royal Observatory, Edinburgh EH9 3HJ, United Kingdom
    \and
    Oukaimeden Observatory, High Energy Physics and Astrophysics Laboratory, Cadi Ayyad University, Morocco
    }
\date{Received ; accepted: }
\titlerunning{FeI and NiI in cometary atmospheres} 
\authorrunning{D. Hutsem\'ekers et al.}
\abstract{FeI and NiI emission lines have  recently been found in the spectra of 17 Solar System comets observed at heliocentric distances between 0.68 and 3.25~au and in the interstellar comet 2I/Borisov. The blackbody equilibrium temperature at the nucleus surface is too low to vaporize the refractory dust grains that contain metals, making the presence of iron and nickel atoms in cometary atmospheres a puzzling observation. Moreover, the measured NiI/FeI abundance ratio is on average one order of magnitude larger than the solar photosphere value. We report new measurements of FeI and NiI production rates and abundance ratios for the Jupiter-family comet (JFC)  46P/Wirtanen in its 2018 apparition and from archival data of the Oort-cloud comet (OCC) C/1996 B2 (Hyakutake). The comets were at geocentric distances of 0.09~au and 0.11~au, respectively. The emission line surface brightness was found to be inversely proportional to the projected distance to the nucleus, confirming that FeI and NiI atoms are ejected from the surface of the nucleus or originate from a short-lived parent. Considering the full sample of 20 comets, we find that the range of NiI/FeI abundance ratios is significantly larger in JFCs than in OCCs. We also unveil significant correlations between NiI/FeI and C$_2$/CN, C$_2$H$_6$/H$_2$O, and NH/CN. Carbon-chain- and NH-depleted comets show the highest NiI/FeI ratios. The existence of such relations suggests that the diversity of NiI/FeI abundance ratios in comets could be related to the cometary formation rather than to subsequent processes~in~the~coma.
}
\maketitle
%
%
%

\section{Introduction}
\label{sec:intro}

\begin{table*}[t]
\caption{Observing circumstances.}
\label{tab:obs}
\centering
\begin{tabular}{lccccccccc}
\hline\hline
 Comet &  Type & Date & $r_h$ &  $\dot{r_h}$ & $\Delta$ & $\dot{\Delta}$ & $d$ & $w$ & $h$ \\
       &       & yyyy-mm-dd & au  &  km s$^{-1}$      & au       & km s$^{-1}$        & \arcsec & \arcsec & \arcsec \\
\hline 
46P/Wirtanen    & JFC & 2018-12-09   & 1.06  & $-$1.2 & 0.09 & $-$5.3 &  0 &  0.44 &  9.5   \\
C/1996 B2 (Hyakutake) & EXT & 1996-03-26 & 1.02  & $-$36.7 & 0.11 & 18.3-20.8 &  0 &  0.87 &  7.4   \\
C/1996 B2 (Hyakutake) & EXT & 1996-03-26 & 1.02  & $-$36.7 & 0.11 & 18.3-20.8 &  2 &  0.87 &  7.4   \\
\hline
\end{tabular}
\tablefoot{$r_h$ and $\Delta$ are the heliocentric and geocentric distances. $\dot{r_h}$ and $\dot{\Delta}$ are the corresponding velocities. $d$, $w$, and $h$ refer to the offset of the slit with respect to the photocenter, the slit width, and the slit height, respectively. Type refers to the comet's dynamical class.}
\end{table*}

\begin{table*}[t]
\caption{FeI and NiI production rates.}
\label{tab:data}
\centering
\begin{tabular}{lccccccccc}
\hline\hline
 Comet & n$_{\rm lines}$  & log$_{10}$ Q(FeI) & log$_{10}$ Q(NiI) & log$_{10}$ [Q(NiI)/Q(FeI)] \\
       &  FeI / NiI  &   s$^{-1}$     &   s$^{-1}$ &             \\
\hline 
46P/Wirtanen          &  6 / 5   &  21.18 $\pm$ 0.12  &  20.80 $\pm$ 0.09  &  $-$0.38  $\pm$  0.15 \\
C/1996 B2 (Hyakutake) & 18 / 12  &  22.91 $\pm$ 0.09  &  22.83 $\pm$ 0.06  &  $-$0.07  $\pm$  0.11 \\  
C/1996 B2 (Hyakutake) & 14 / 12  &  22.97 $\pm$ 0.05  &  22.84 $\pm$ 0.06  &  $-$0.14  $\pm$  0.08 \\ 
\hline
\end{tabular}
\end{table*}

Metals, in particular iron and nickel, have been found in cometary dust by in situ experiments on board the Giotto and Rosetta spacecrafts~\citep{Jessberger1988,Stenzel2017} as well as in dust particles collected by the Stardust space probe~\citep{Zolensky2006,Gainsforth2019}. They essentially appear in silicate, sulfide, and metal grains. Two Sun-grazing comets, C/1882 R1, the Great Comet of 1882 \citep{Copeland1882}, and C/1965 S1 (Ikeya-Seki)~\citep{Dufay1965,Curtis1966,Thackeray1966,Preston1967,Slaughter1969}, approached the Sun so close -- at perihelion distances of $\simeq$~0.01 au -- that dust grains vaporized, revealing lines of several metals in the coma spectrum, including iron, nickel, chromium, manganese, and cobalt (the iron-group elements). In comet Ikeya-Seki, FeI lines were observed pre-perihelion at heliocentric distances $r_h =$ 0.09 to 0.05 au (NiI lines were out of the observed spectral range), while lines of all iron-group elements were observed post-perihelion at $r_h =$ 0.04 to 0.14 au. About 80 FeI and NiI emission lines were identified in the ultraviolet spectrum of comet Ikeya-Seki, from which a NiI/FeI abundance ratio comparable to that of chondrites and the solar photosphere was derived \citep{Preston1967,Arpigny1978,Arpigny1979,Manfroid2021}. The presence of iron vapor has also been claimed in the bright comet C/2006 P1 (McNaught) based not on spectroscopy but on the dynamical properties of a faint tail observed at perihelion ($r_h$ = 0.17 to 0.19 au) by the Solar Terrestrial Relations Observatory (STEREO) spacecraft \citep{Fulle2007}. In this case, the iron atoms were thought to sublimate from sulfide grains (e.g., tro{\"i}lite). Fe$^{+}$ ions were also tentatively identified in the coma of comet 1P/Halley at $r_h \simeq$ 0.9~au \citep{Krankowsky1986,Ibadov1992}.

Over the last 20 years, dozens of FeI and NiI emission lines have been found in the 3000-4000~\AA\ spectral region of 17 comets at heliocentric distances between 0.68 and 3.25~au, observed with the high-resolution Ultraviolet-Visual Echelle Spectrograph (UVES) mounted on the European Southern Observatory (ESO) Very Large Telescope \citep[VLT;][]{Manfroid2021}. Finding numerous FeI and NiI emission lines in the spectra of comets at heliocentric distances of up to 3~au, where the surface blackbody equilibrium temperature $T \simeq 280 \, r_h^{-1/2}$ K is far too low to allow sublimation of silicates ($T_{\rm sub} \gtrsim $ 1200~K) and sulfides ($T_{\rm sub} \gtrsim$ 600~K), came as a surprise. Moreover, the average abundance ratio of all measurements, log$_{10}$(NiI/FeI) = $-0.06 \pm 0.31$, differs by one order of magnitude from the ratio log$_{10}$(Ni/Fe) = $-1.10 \pm 0.23$ estimated in the dust of 1P/Halley \citep{Jessberger1988} and the log$_{10}$(NiI/FeI) = $-1.11 \pm 0.09$ measured in the coma of the Sun-grazing comet Ikeya-Seki using the fluorescence model of \citet{Manfroid2021}. No evidence of other metal lines (except sodium) were found in the UVES spectra; notable was the lack of chromium, the next most abundant iron-group element in the Sun after nickel.

\citet{Manfroid2021} proposed several mechanisms to explain these observations and, more specifically, how FeI and NiI atoms can be released at such low cometary temperatures and why the NiI/FeI abundance ratio is one order of magnitude higher than the solar value. In particular, the superheating of Ni-rich sulfides and the sublimation of organometallic complexes such as carbonyls were discussed. However, interpretations remain open, requiring more observational constraints.

In this paper we analyze new high-resolution observations of the Jupiter-family comet (JFC) 46P/Wirtanen together with archived spectra of the Oort-cloud comet (OCC) C/1996 B2 (Hyakutake), ``the great comet of 1996,'' focusing on the FeI and NiI emission lines. The spectra were recorded when the comets were closer to Earth than any other comet in our sample. Observations, data reduction, and measurements are described in Sect.~\ref{sec:obs}. Results that unveil differences between JFCs and OCCs are described in Sect.~\ref{sec:results}. Discussion and conclusions are given in Sects.~\ref{sec:discu} and~\ref{sec:conclu}.

Throughout the paper, the comets are classified according to their dynamical class \citep{Levison1996}: JFC corresponds to ecliptic comets with short periods ($<$ 200 years),  Halley-type (HT) corresponds to nearly isotropic comets with a semimajor axis $a <$ 40~au, external (EXT) corresponds to nearly isotropic comets with 40~au $< a <$ 10000~au, and NEW corresponds to comets that come directly from the Oort cloud ($a >$ 10000~au). To emphasize the difference with JFCs, we also refer to the  nearly isotropic NEW, EXT, and HT comets as OCCs. 

\begin{figure}[t]
\centering
\resizebox{\hsize}{!}{\includegraphics*{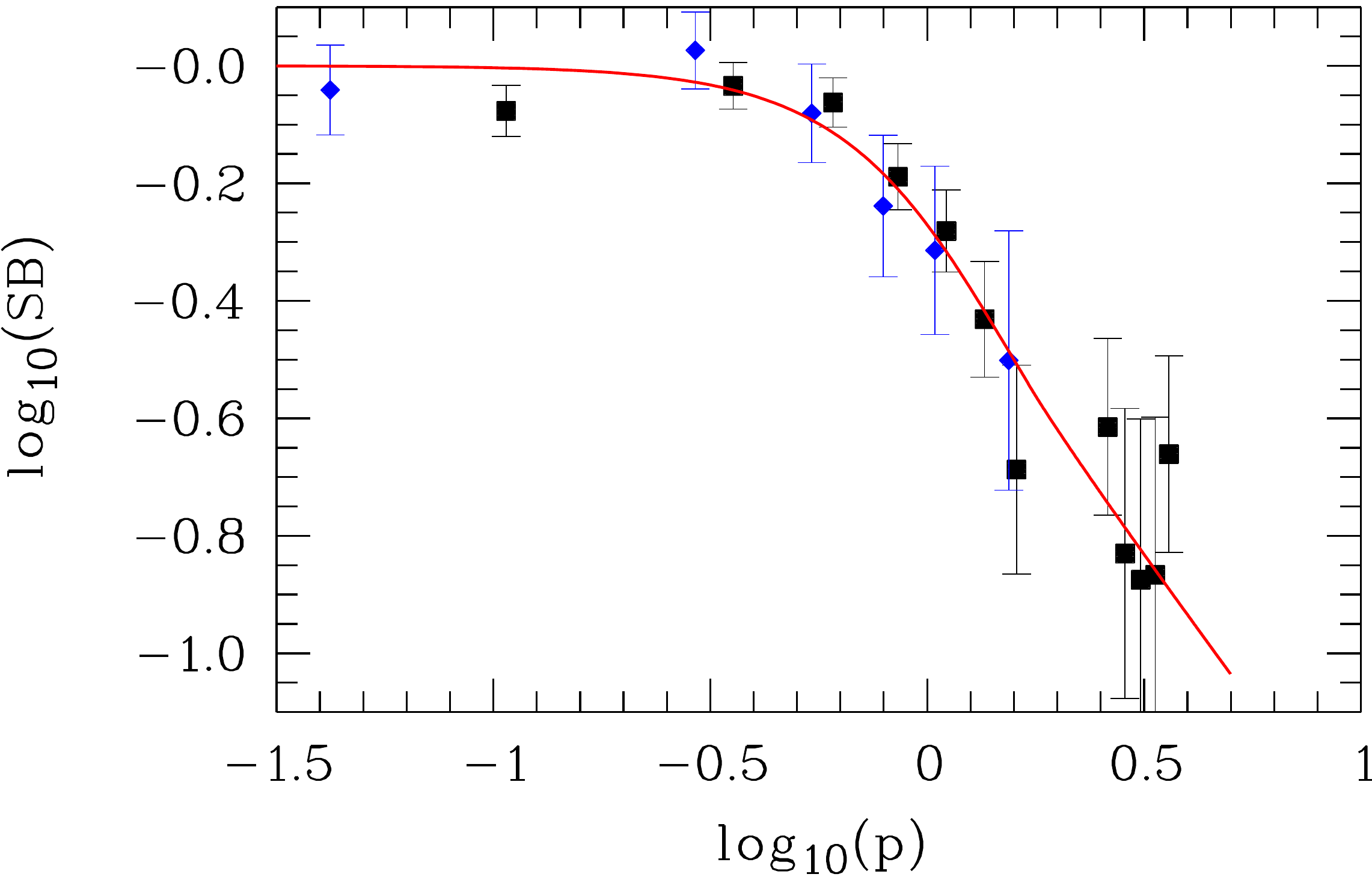}}
\caption{Spatial profile of the brightest FeI line ($\lambda$3719~\AA ; black squares) and the brightest NiI line  ($\lambda$3458~\AA ; blue diamonds) observed in comet 46P. The measured surface brightness (SB; in arbitrary units) is plotted as a function of the projected nucleocentric distance $p$ in arcsec. The red line represents a $p^{-1}$ distribution of the SB convolved with a 1.35\arcsec\ FWHM Gaussian that accounts for the seeing and tracking imperfections.}
\label{fig:rp}
\end{figure}

\section{Observations, data reduction, and measurements}
\label{sec:obs}
Comet 46P was observed at the ESO VLT with the UVES spectrograph on December 9, 2018. Two different settings were used so as to cover the spectral range 3050-10400\AA\ with a few gaps. Two exposures were obtained for each setting. The slit width was 0\farcs44, providing a resolving power R~$\simeq$~80000 over the full spectral range. The slit length was 9.5\arcsec\ for the blue settings that contain the FeI and NiI lines of interest. The observing circumstances are given in Table~\ref{tab:obs}. The data were reduced with the UVES pipeline with custom procedures for spectrum extraction and cosmic ray removal. The scattered spectrum of the Sun (dust, twilight) was removed as described in \citet{Manfroid2009}. The two exposures were finally averaged. 

The spectra of comet Hyakutake were obtained at the Kitt Peak National Observatory using the Echelle Spectrograph on the 4m Mayall Telescope on March 26, 1996. With a 0\farcs87 $\times$ 7\farcs4 slit, the echelle spectrograph had a resolving power R~$\simeq$~18000. Four offset positions of the slit with respect to the photocenter were used. The archived extracted and averaged spectra were retrieved from the NASA Planetary Data System \citep{AHearn2015}. A detailed description of the observations and data reduction is provided in \citet{Meier1998}. Only the two spectra secured with 0\arcsec\ and 2\arcsec\ slit offsets are considered here. Spectra obtained at larger offsets contain less useful lines, and the offset positions are less accurate. The observing circumstances are also summarized in Table~\ref{tab:obs}.

Examples of FeI and NiI lines are shown in Figs.~\ref{fig:spec1} and~\ref{fig:spec2} for comets Hyakutake and 46P, respectively. In comet Hyakutake, the FeI and NiI lines are bright and some of them were reported as unidentified features by \citet{Kim2003}. The FeI and NiI lines are much fainter in the spectrum of comet 46P, though they are clearly detected thanks to the high spectral resolution provided by UVES.

The intensities of unblended FeI and NiI emission lines were then measured in the three spectra and compared to the intensities computed using a dedicated fluorescence model in order to derive the production rates. The procedure is described in detail in \citet{Manfroid2021}. The resulting production rates, Q, and the derived abundance ratios are given in Table~\ref{tab:data}, together with the number of lines used in the analysis. They are consistent with the values reported in \citet{Manfroid2021} for other comets\footnote{\citet{Bromley2021} independently measured the NiI/FeI abundance ratio in comet Hyakutake and found a value that is also consistent with the abundance ratio measured in other comets.}.  Production rates of other molecular species, in particular CN, C$_2$, and NH,  were also derived from the UVES spectra as done in \citet{Manfroid2021}. The measured C$_2$/CN and NH/CN abundance ratios are given in Table~\ref{tab:data2}.

\section{Analysis and results}
\label{sec:results}

\begin{figure}[t]
\centering
\resizebox{\hsize}{!}{\includegraphics*{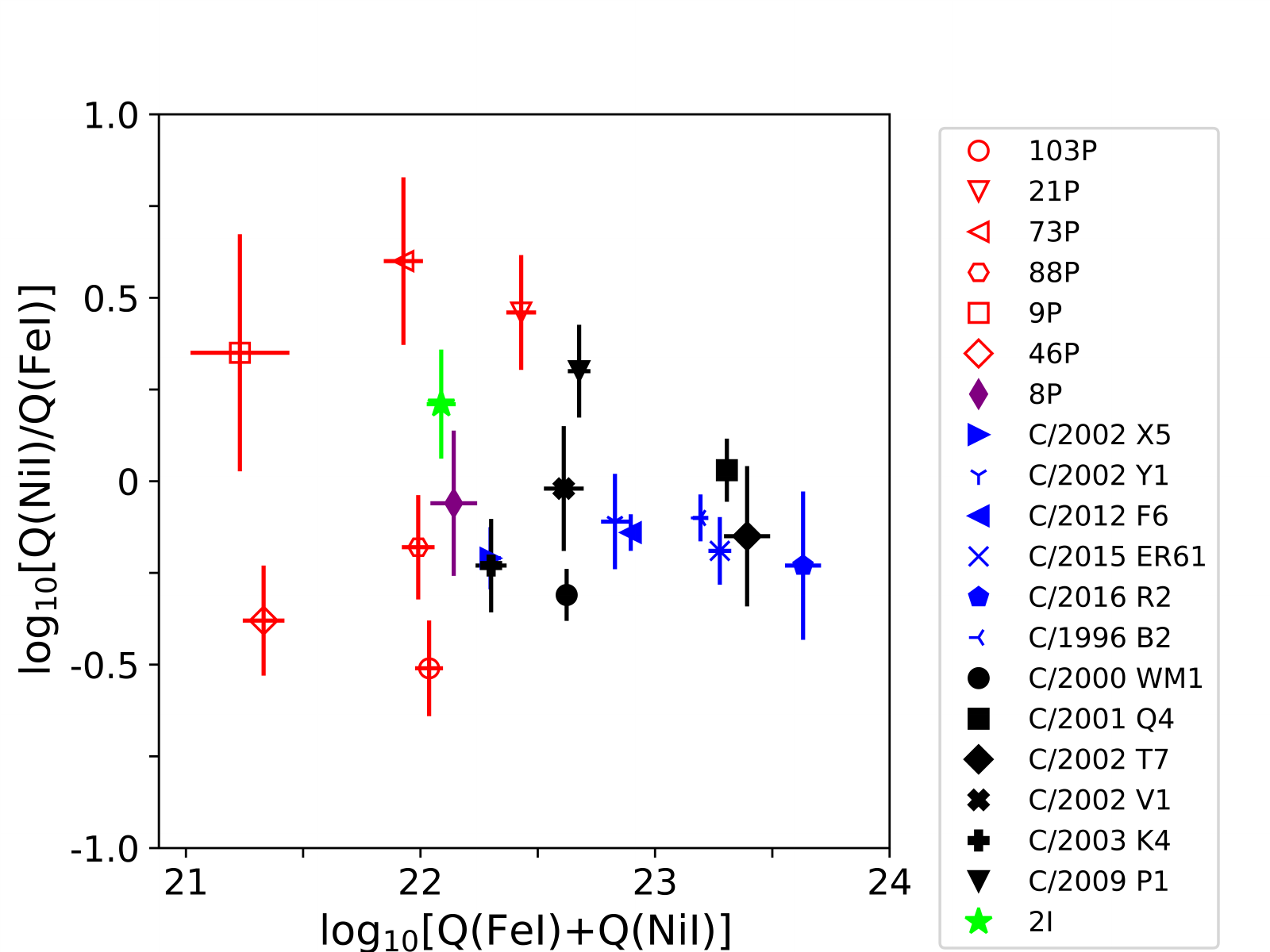}}
\caption{NiI/FeI abundance ratio against the total FeI+NiI production rate (in s$^{-1}$). Colors represent the comet types: JFC (red), EXT (blue), NEW (black), HT (violet), and interstellar (green).}
\label{fig:cor1}
\end{figure}

One of the major characteristics of the FeI and NiI emission lines is their short spatial extent. In the spectra obtained for comet 103P/Hartley~2 at a geocentric distance $\Delta =$ 0.17~au, the surface brightness was found to be inversely proportional to the projected distance to the nucleus \citep{Manfroid2021}. Such a spatial profile indicates that the FeI and NiI atoms are ejected from the surface of the nucleus or originate from a short-lived parent. Although the signal-to-noise is not as good for comet 46P as for comet 103P/Hartley~2, we see in Fig.~\ref{fig:rp} that the $p^{-1}$ distribution also reproduces the spatial profile observed in comet 46P during a much closer encounter at $\Delta =$ 0.09~au. At this distance, 1\arcsec\ represents 65~km in projection on the coma. FeI and NiI atoms should thus originate at nucleocentric distances $\lesssim$ 50~km, assuming a blurring of 1.35\arcsec\ full width at half maximum (FWHM) that accounts for the seeing and tracking imperfections.

In Fig.~\ref{fig:cor1} we show the NiI/FeI abundance ratio as a function of the total FeI plus NiI production rate for the 20 comets studied so far: 17 comets from \citet{Manfroid2021}, the interstellar comet 2I/Borisov from \citet{Opitom2021}, and 46P and Hyakutake (this work).  When more than one measurement is available for a given comet, we used the simple average value with an error computed from the individual measurement errors. Jupiter-family comets clearly cluster at low production rates. This characteristic of JFCs was also found for other species and is indicative of their lower activity \citep{AHearn1995,Moulane2021}. In the current sample, the NiI/FeI abundance ratios averaged for the different cometary types do not differ within the errors, as already noticed in \citet{Manfroid2021}. However, it is clear from  Fig.~\ref{fig:cor1} that JFCs show a much larger range of NiI/FeI ratios than OCCs. An F-test of variance indicates that the probability that the NiI/FeI ratio has the same variance in the JFC and OCC samples is 0.2\%. Most of the difference in variance comes from the difference between JFCs and EXT comets, with NEW comets showing an intermediate dispersion.

\begin{figure}[t]
\centering
\resizebox{\hsize}{!}{\includegraphics*{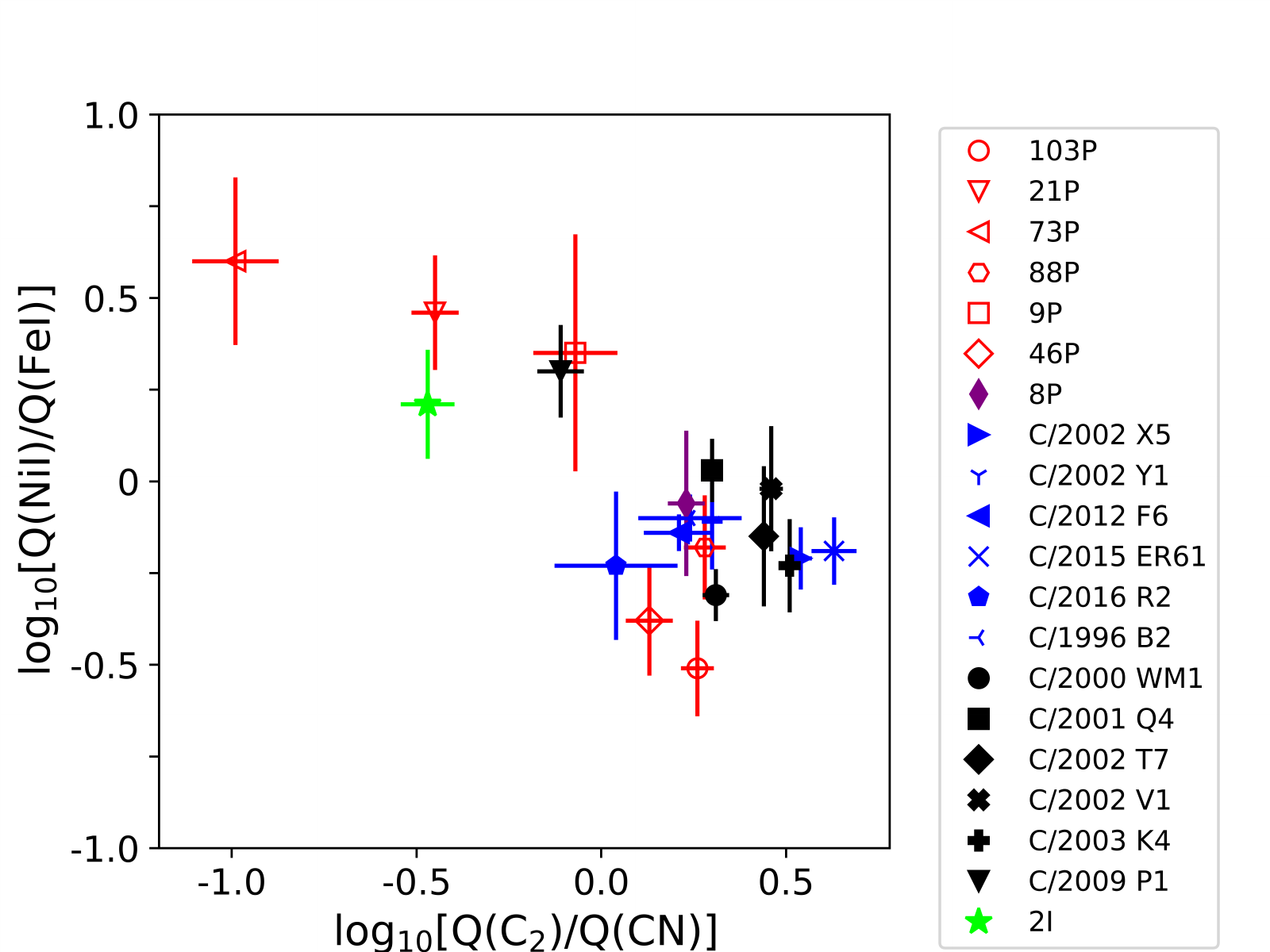}}
\caption{NiI/FeI abundance ratio against the C$_2$/CN ratio. Colors represent the comet types as in Fig.~\ref{fig:cor1}. }
\label{fig:cor2}
\end{figure}

\begin{figure}[t]
\centering
\resizebox{\hsize}{!}{\includegraphics*{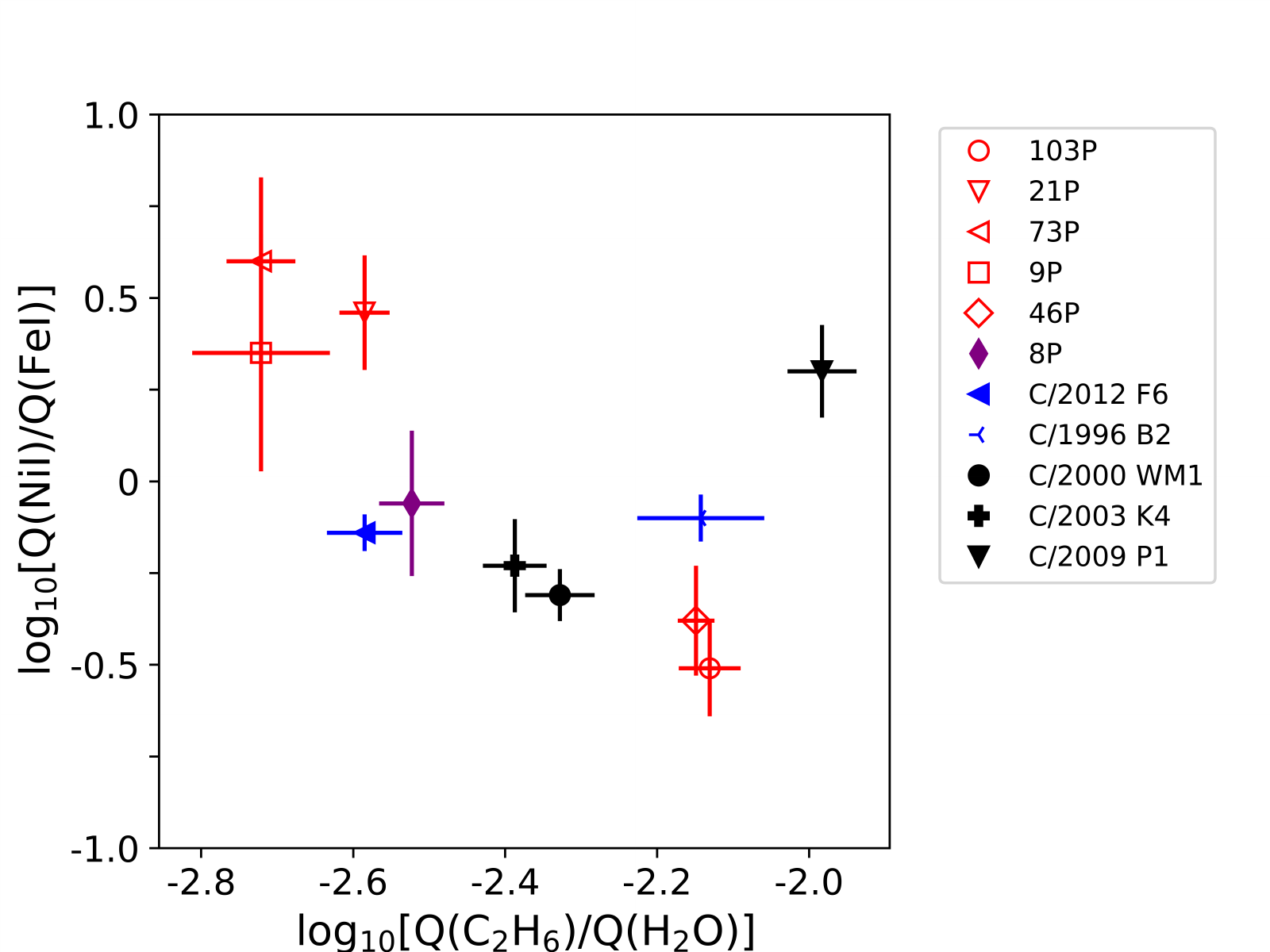}}
\caption{NiI/FeI abundance ratio against the C$_2$H$_6$/H$_2$O ratio.  The C$_2$H$_6$/H$_2$O ratios are from \citet{DelloRusso2016} except for the C$_2$H$_6$/H$_2$O ratios for comets 21P/Giacobini-Zinner and 46P, which are from \citet{Roth2020,Roth2021}. Colors represent the comet types as in Fig.~\ref{fig:cor1}.}
\label{fig:cor3}
\end{figure}

Jupiter-family comets are known to show a larger dispersion of some observables, in particular the C$_2$/CN abundance ratio that is used to characterize carbon-chain-depleted comets \citep{AHearn1995,Schleicher2008,Cochran2012,Moulane2021}. In Fig.~\ref{fig:cor2} we show the NiI/FeI abundance ratio as a function of the C$_2$/CN abundance ratio measured in our spectra (Table~\ref{tab:data2}). Our sample includes the well-known carbon-chain-depleted comets 21P/Giacobini-Zinner \citep{Schleicher1987,Moulane2021} and 73P/Schwassmann-Wachmann \citep{Schleicher2011}, as well as the very peculiar comet  C/2016 R2 \citep[PANSTARRS;][]{Cochran2018,Biver2018,Opitom2019} and the interstellar comet 2I/Borisov \citep{Fitzsimmons2019}. A correlation is clearly seen, in the sense that comets with low C$_2$/CN ratios have high NiI/FeI values. Kendall and Spearman rank-order (K+S) tests give a probability of 1.5\% that the two quantities are not correlated. The correlation is mostly driven by the large range of values in the JFC sample, but it is also observed when comparing comets C/2009 P1 (Garradd) and 2I/Borisov to other OCCs, both comets showing a smaller C$_2$/CN ratio and a higher NiI/FeI ratio.

The highly volatile species C$_2$H$_2$ and C$_2$H$_6$ that are observed in the near-infrared and likely related to the production of C$_2$ are also often depleted in JFCs \citep{DelloRusso2016}. We found 11 measurements of C$_2$H$_6$ in the literature for the comets of our sample\footnote{Only seven C$_2$H$_2$ measurements were found. Moreover, they are affected by large uncertainties, making this data set too poor to be useful.}. When several measurements are available for a comet, we adopted the value that is closest in time to the UVES observations. In Fig.~\ref{fig:cor3} we show the NiI/FeI ratio against the C$_2$H$_6$/H$_2$O ratio. The high dispersion of JFCs is again clearly seen, as is a correlation between the NiI/FeI and C$_2$H$_6$/H$_2$O abundance ratios. K+S tests give a probability of 4\% that the two quantities are not correlated. It should be emphasized that the UVES and infrared observations are most often not contemporaneous. With respect to the global trend, comet C/2009 P1 (Garradd) appears as an outlier; however, if the water production rate of this comet was actually underestimated by a factor of two to three, as suggested by \citet{Paganini2012} and \citet{Combi2013}, C/2009 P1 (Garradd) would also follow the trend and the correlation would be much more significant (probability of 0.4\% that the two quantities are not correlated). On the other hand, only JFCs might be involved in the correlation. More simultaneous optical and infrared data sets are necessary to clarify this issue.

\begin{figure}[t]
\centering
\resizebox{\hsize}{!}{\includegraphics*{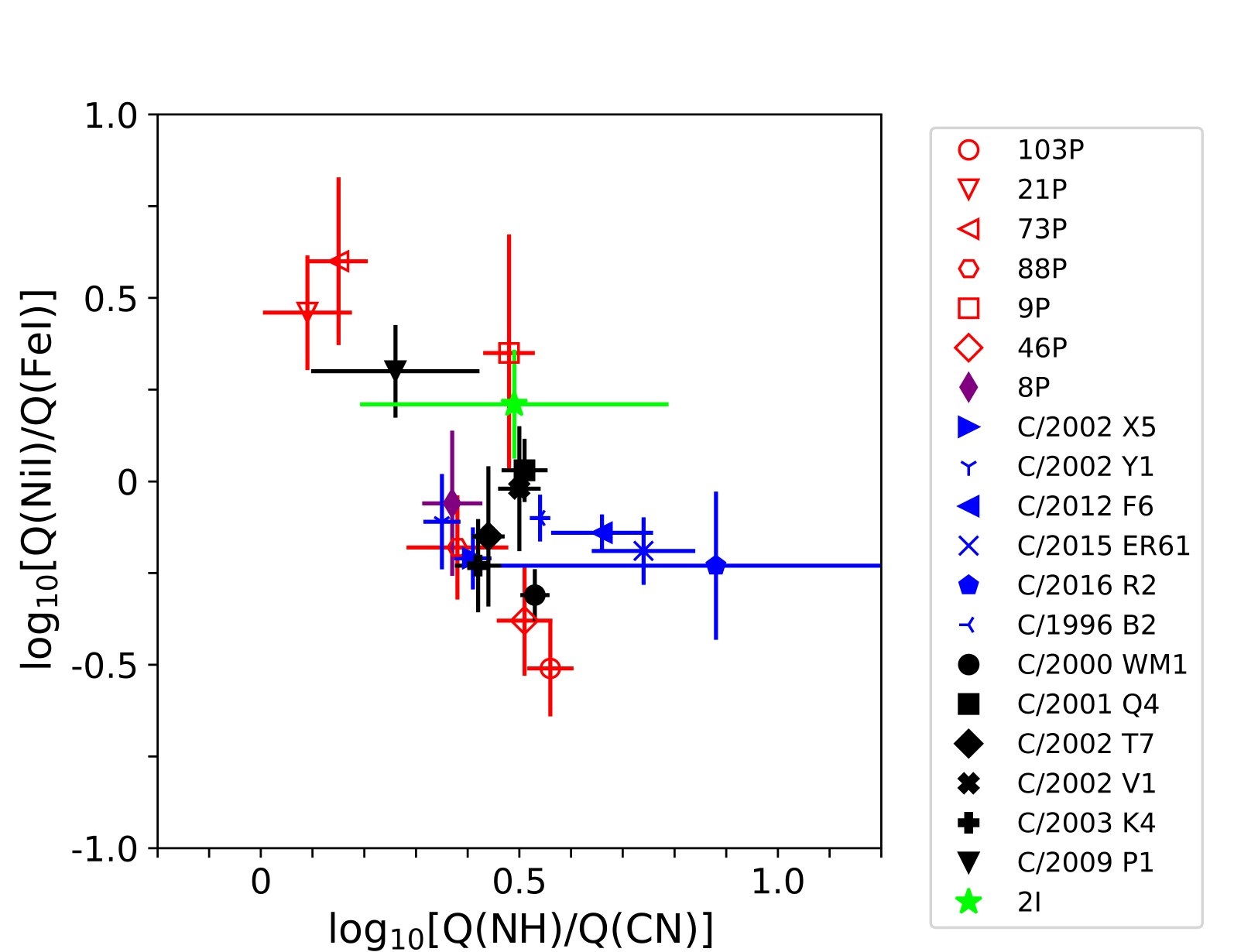}}
\caption{NiI/FeI abundance ratio against the NH/CN ratio. Colors represent the comet types as in Fig.~\ref{fig:cor1}. }
\label{fig:cor4}
\end{figure}

Since there is evidence that some carbon-chain JFCs are also depleted in NH2 and NH \citep{Fink2009,Cochran2012,Pierce2021}, we looked for a correlation between the NiI/FeI and NH/CN abundance ratios measured in our spectra (Table~\ref{tab:data2}). Figure~\ref{fig:cor4} shows this correlation, the comets with the smallest NH/CN ratios showing the highest NiI/FeI ratios. K+S tests give a probability of 1\% that the two quantities are not correlated. While dominated by JFCs, the correlation also involves C/2009 P1 (Garradd), which has low NH/CN and high NiI/FeI ratios. It should be noted that \citet{Pierce2021} found a tentative correlation between NH2/CN and C$_2$/CN in JFCs that might also be present between NH/CN and C$_2$/CN (Fig.~\ref{fig:cor9}).

\section{Discussion}
\label{sec:discu}

Various surveys have established the distinction between carbon-chain-depleted and typical comets as one of the most robust compositional classifications. Furthermore, they provided evidence that carbon-chain depletion reflects the primordial composition at the location of cometary formation rather than subsequent evolution \citep{AHearn1995,Fink2009,Schleicher2008,Cochran2012,Schleicher2014,Moulane2021}. One of the strongest arguments that carbon-chain depletion is related to cometary formation rather than to evolution comes from the identical depletion measured in the various pieces of the split comet 73P/Schwassmann-Wachmann~3 \citep{Schleicher2011}. The fact that the NiI/FeI abundance ratio appears correlated with the C$_2$/CN and the C$_2$H$_6$/H$_2$O ratios thus suggests that the range of NiI/FeI values observed in our comet sample is also related to their primordial composition rather than to evolutionary effects. This also means that the NiI/FeI ratio is not entirely related to the different sublimation rates of the species releasing the NiI and FeI atoms, a hypothesis discussed in \citet{Manfroid2021}. While primarily observed in JFCs, atypical NiI/FeI abundance ratios are also observed in some OCCs, a fact well established for the C$_2$/CN abundance ratio that suggests that the reservoirs at the origin of JFCs and OCCs were mixed at some epoch \citep[e.g.,][]{Cochran2020}.  Interestingly, the depletion of simple organics such as C$_2$H$_6$ could be related to their conversion into more complex organics \citep{Roth2020,Ootsubo2020}. This is in agreement with the unusual polarization properties observed in comets 21P/Giacobini-Zinner, 73P/Schwassmann-Wachmann~3, and 9P/Tempel~1 that were interpreted in terms of a high number of large particles \citep{Kiselev2000,Kiselev2015}. Complex organics might contain metals, such as the organometallic compounds discussed in \citet{Manfroid2021}, and thus possibly pave the way to an explanation of the observed NiI/FeI abundance ratio. 

\section{Conclusions}
\label{sec:conclu}

We have analyzed new and archived high-resolution spectroscopic observations of two comets that approached the Earth at about 0.1~au. We focused on the FeI and NiI emission lines, only recently discovered in cometary atmospheres. We show that, at this spatial resolution, the surface brightness profile of the lines is still inversely proportional to the projected distance to the nucleus, confirming that FeI and NiI atoms are ejected from the surface of the nucleus or originate from a short-lived parent.

We computed the FeI and NiI productions rates and the corresponding NiI/FeI abundance ratios. Considering the full sample of 20 comets, we find that the range of NiI/FeI ratios is significantly larger in JFCs than in OCCs, which is clearly reminiscent of the behavior of the C$_2$/CN abundance ratio.  We unveil significant correlations between NiI/FeI and C$_2$/CN, C$_2$H$_6$/H$_2$O, and NH/CN. Carbon-chain- and NH-depleted comets show the highest NiI/FeI ratios. The existence of such a relation suggests that the diversity of NiI/FeI ratios in comets may be more related to the cometary formation than to subsequent processes in the coma. It also indicates that metal abundances could constitute a new important probe of comet formation and evolution.

\begin{acknowledgements}
We thank the referee, Anita Cochran, and the editor, Emmanuel Lellouch, for useful comments on the manuscript. We thank Dave Schleicher for sharing his photometric production rates for comets in common with our UVES sample to cross-check our spectroscopic C$_2$/CN ratios. JM, DH, and EJ are honorary Research Director, Research Director and Senior Research Associate at the F.R.S-FNRS, respectively. CO is a Royal Astronomical Society Norman Lockyer fellow and University of Edinburgh Chancellor's fellow.
\end{acknowledgements}

\bibliographystyle{aa}
\bibliography{references}

\begin{appendix}
\onecolumn
\section{Supplementary data}
\label{sec:appendix}

\begin{figure*}[h]
\centering
\resizebox{14cm}{!}{\includegraphics*{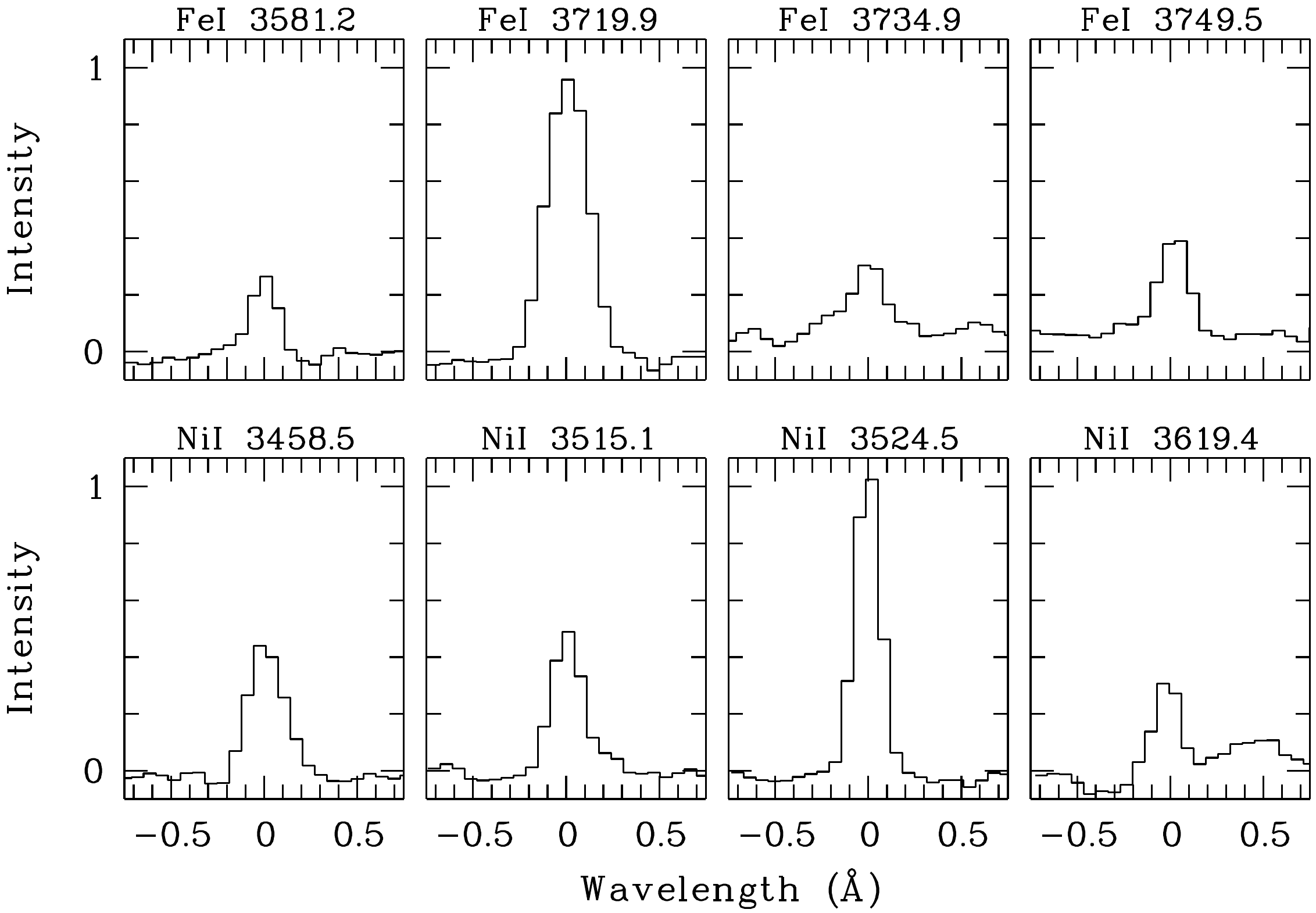}}
\caption{Examples of FeI and NiI lines detected in the spectrum of comet C/1996 B2 (Hyakutake). Each sub-image covers a spectral range of 1.5~\AA,\ centered on the line. The intensity is given in arbitrary units.}
\label{fig:spec1}
\end{figure*}
\begin{figure*}[h]
\centering
\resizebox{14cm}{!}{\includegraphics*{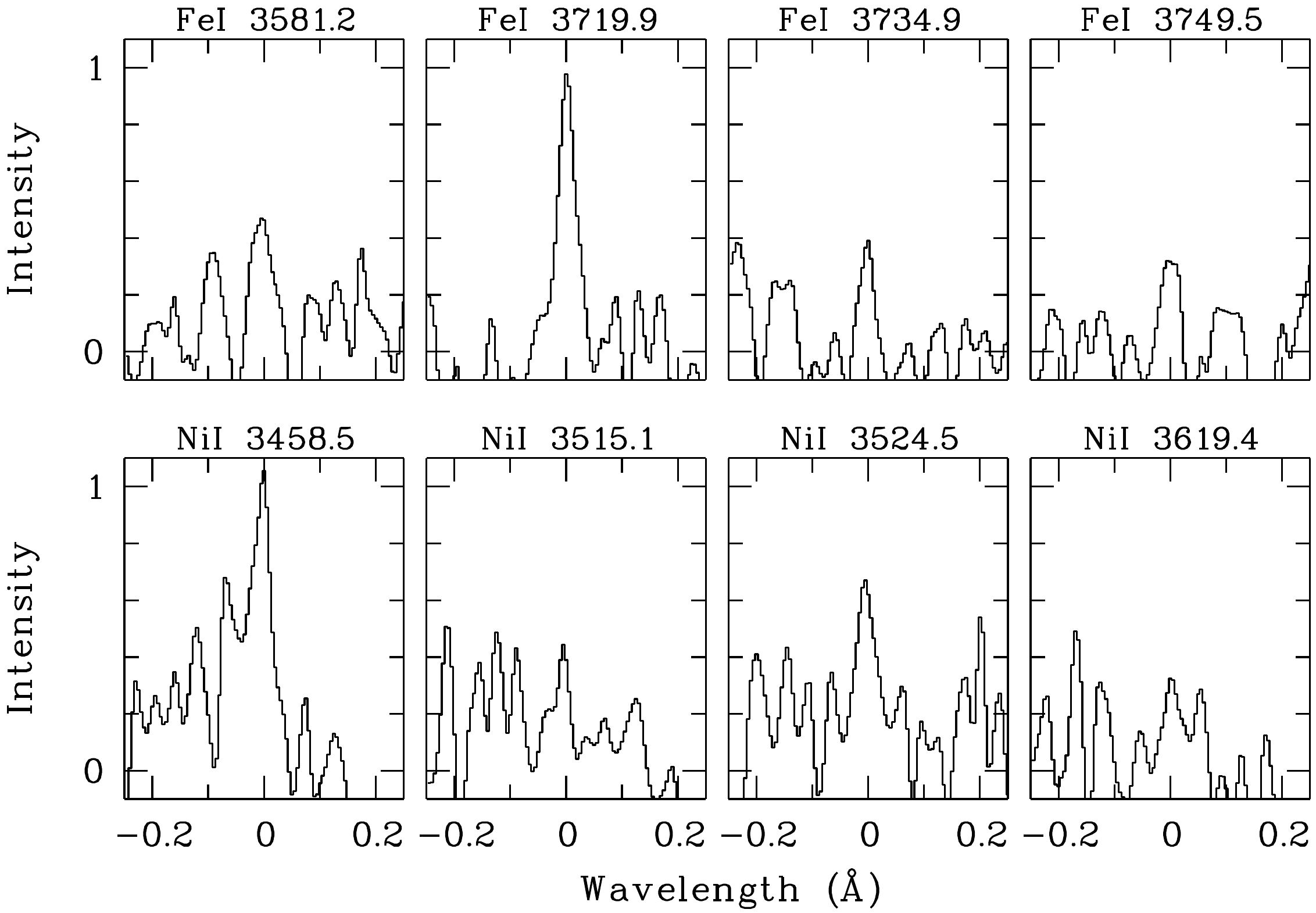}}
\caption{Examples of FeI and NiI lines detected in the spectrum of comet 46P/Wirtanen. Each sub-image covers a spectral range of 0.5~\AA,\ centered on the line. The intensity is given in arbitrary units.}
\label{fig:spec2}
\end{figure*}

$ $ 

\begin{table*}
\caption{C$_2$/CN and NH/CN abundance ratios.}
\label{tab:data2}
\centering
\begin{tabular}{lrr}
\hline\hline
 Comet & log [Q(C$_2$)/Q(CN)] & log [Q(NH)/Q(CN)] \\
\hline 
103P/Hartley 2              &     0.26$\pm$0.04  &   0.56$\pm$0.04    \\
21P/Giacobini-Zinner        &  $-$0.45$\pm$0.06  &   0.09$\pm$0.09    \\
73P/Schwassmann-Wachmann 3  &  $-$0.99$\pm$0.12  &   0.15$\pm$0.06    \\
88P/Howell                  &     0.28$\pm$0.06  &   0.38$\pm$0.10    \\
9P/Tempel 1                 &  $-$0.07$\pm$0.11  &   0.48$\pm$0.05    \\
46P/Wirtanen                &     0.13$\pm$0.06  &   0.51$\pm$0.05    \\
8P/Tuttle                   &     0.23$\pm$0.05  &   0.37$\pm$0.06    \\
C/2002 X5 (Kudo-Fujikawa)   &     0.54$\pm$0.03  &   0.41$\pm$0.04    \\
C/2002 Y1 (Juels-Holvorcem) &     0.30$\pm$0.03  &   0.35$\pm$0.04    \\
C/2012 F6 (Lemmon)          &     0.21$\pm$0.09  &   0.66$\pm$0.10    \\
C/2015 ER61 (PANSTARRS)     &     0.63$\pm$0.06  &   0.74$\pm$0.10    \\
C/2016 R2 (PANSTARRS)       &     0.04$\pm$0.17  &   0.88$\pm$0.48    \\
C/1996 B2 (Hyakutake)       &{\it 0.24$\pm$0.14} &   0.54$\pm$0.02    \\
C/2000 WM1 (LINEAR)         &     0.31$\pm$0.04  &   0.53$\pm$0.03    \\
C/2001 Q4 (NEAT)            &     0.30$\pm$0.02  &   0.51$\pm$0.04    \\
C/2002 T7 (LINEAR)          &     0.44$\pm$0.01  &   0.44$\pm$0.03    \\
C/2002 V1 (NEAT)            &     0.46$\pm$0.03  &   0.50$\pm$0.04    \\
C/2003 K4 (LINEAR)          &     0.51$\pm$0.02  &   0.42$\pm$0.04    \\
C/2009 P1 (Garradd)         &  $-$0.11$\pm$0.06  &   0.26$\pm$0.16    \\
2I/Borisov                  &  $-$0.47$\pm$0.07  &   0.49$\pm$0.30    \\
\hline
\end{tabular}
\tablefoot{The C$_2$/CN ratio of comet Hyakutake is from \citet{Schleicher2002}.}
\end{table*}

$ $ 

\begin{figure*}
\centering
\resizebox{10cm}{!}{\includegraphics*{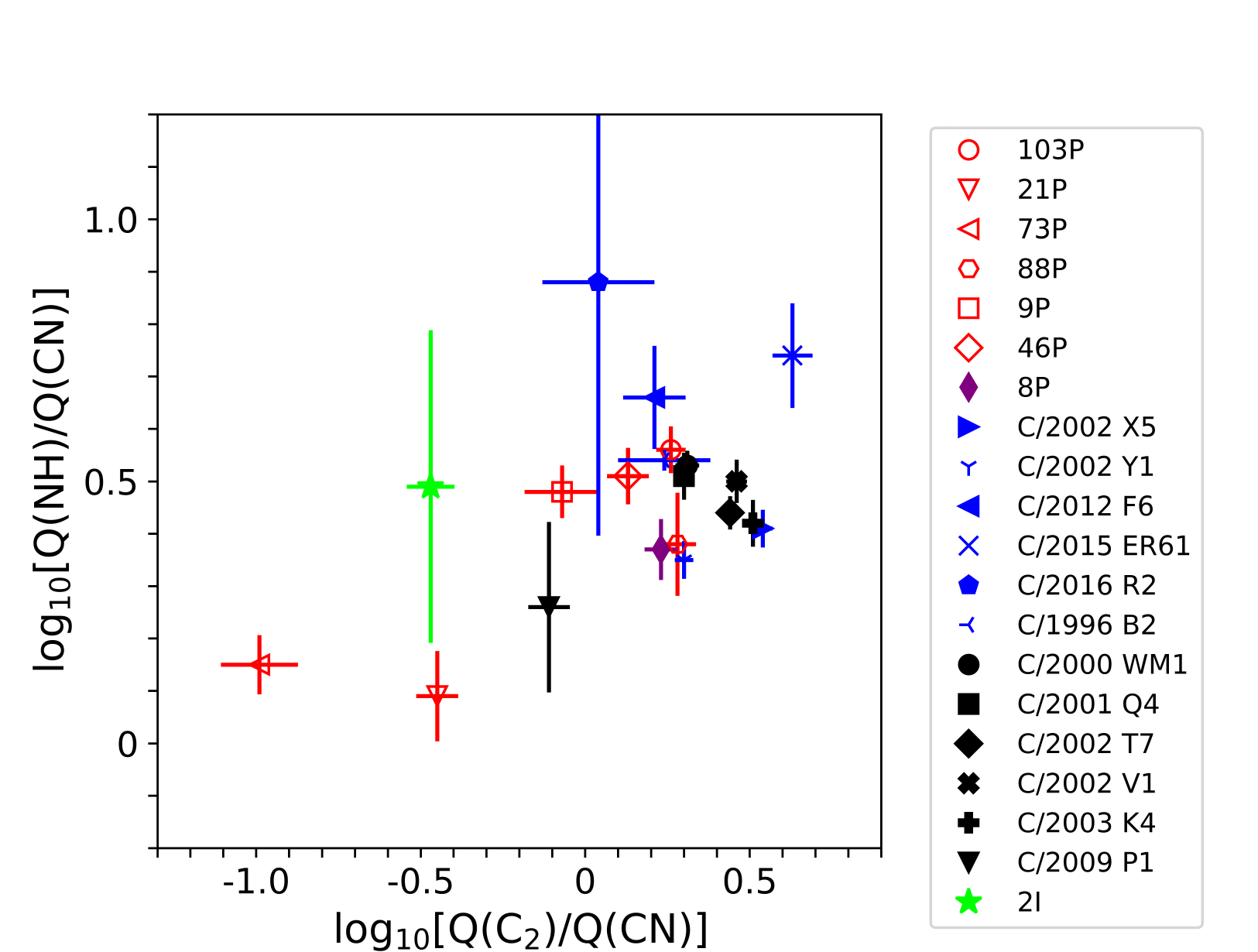}}
\caption{NH/CN abundance ratio against the C$_2$/CN ratio. Colors represent the comet types as in Fig.~\ref{fig:cor1}.}
\label{fig:cor9}
\end{figure*}

\end{appendix}

\end{document}